\documentclass[11pt]{article}

\usepackage{amsmath}
\usepackage[margin=1in]{geometry}
\usepackage{listings}

\title{Optimizing Secure Statistical Computations with PICCO}

\author{
Justin DeBenedetto\\
University of Notre Dame\\
Notre Dame, IN\\
jdebened@nd.edu
\and
Marina Blanton\\
University at Buffalo\\
Buffalo, NY\\
mblanton@buffalo.edu
}

\date{}

\begin{document}
\maketitle

\begin{abstract}
Growth in research collaboration has caused an increased need for
sharing of data. However, when this data is private, there is also an
increased need for maintaining security and privacy.  Secure
multi-party computation enables any function to be securely evaluated
over private data without revealing any unintended data. A number of
tools and compilers have been recently developed to support evaluation
of various functionalities over private data. PICCO is one of such
compilers that transforms a general-purpose user program into its
secure distributed implementation. Here we assess performance of
common statistical programs using PICCO. Specifically, we focus on
chi-squared and standard deviation computations and optimize user
programs for them to assess performance that an informed user might
expect from securely evaluating these functions using a
general-purpose compiler.
\end{abstract}

\section{Introduction}

Research collaboration among universities continues to increase in prevalence and importance \cite{Adams2012}.  As collaboration grows, so does the need to share data.  However, in certain instances this data can be sensitive.  One such domain is institutional research in which comparative data measures are used for internal institutional assessment as well as by larger accrediting agencies.  The interest in such data assessments is evidenced by the Cooperative Institutional Research Program (CIRP) \cite{eagan2014american}.

In fall 2014, over 150,000 first-time, full-time students from 227 different colleges and universities across the United States participated in the CIRP Freshman Survey.  This collaborative administration of a survey provided data to each college and university on how they compare nationally \cite{eagan2014american}.  This demonstrates the interest of universities in having such data available for analysis, but there are limitations.  Student grades and other sensitive student data, aggregate figures which risk identifying small groups of single students, and other such factors often inhibit the ability to freely share data between interested institutions.

Another domain in which sensitive data sharing continues to grow is
healthcare \cite{yau2008}.  It is often beneficial to have more patient data available for medical research, but there is significant need to protect the sensitive patient data.  In such environments, data analysis requires multiple parties to share private data for computations without allowing any other party involved to obtain the input data.

Secure multi-party computation or secure computation outsourcing can
provide a solution for such data protection in these circumstances.
Such techniques protect privacy of data throughout the computation and
do not reveal any information about private inputs to any party
besides the agreed-upon output. In recent years, a number of compilers
and tools have emerged that enable a user program that handles sensitive data to be compiled into its equivalent implementation that provably protects private data throughout the computation. Of a particular interest to us is the PICCO compiler \cite{picco} that transforms a program written in an extension of C with variables to be protected marked as private into its secure distributed implementation. We thus utilize PICCO for producing secure implementations of statistical computations as a solution for preserving privacy of sensitive data in collaborative research.

The goal of this work is to experiment with PICCO to assess what performance one might expect from secure implementations of statistical programs produced using a general-purpose compiler. In particular, we focus on standard deviation and chi-squared computations. We utilize optimization features available in PICCO and weigh the cost of different operations in the secure computation framework to determine what performance an informed user might expect from secure realizations of statistical functions.

\section{Preliminaries}

We first proceed with describing two statistical tests evaluated in this work in a secure computation setting and describe the tools for securely evaluating the functions.

\subsection{Statistical tests}

\emph{Standard deviation} is a commonly used statistical measure to describe how spread out the data is in a data set.  The standard deviation can be obtained by taking the square root of the variance, or more generally, the square root of the average of squared differences from the mean of the data.  Thus, for a data set $(x_1, \ldots, x_N)$ consisting of $N$ data points the standard deviation is 
\begin{equation} \label{eq:std-dev}
\sigma = \sqrt{\frac{1}{N} \sum_{i=1}^N (x_i - \mu)^2},
\end{equation}
with $\mu$ being the mean of the data set.

\emph{Chi-squared tests} are a popular statistical measure used to test whether the observed frequencies are significantly different than the expected frequencies in any of the data categories. In particular, let the overall data population be divided into $n$ categories according to one variable and into $m$ categories according to another variable. This gives rise to a matrix representation of the data with the value at location $i,j$ corresponding to the number of instances among the population taking the $i$th value of the first variable and the $j$th value of the second variable. Then the chi-squared value is obtained by first computing the sum over all of the cells of the squared difference between observed and expected values divided by the expected value.  That is, this sum can be written as:
$$\sum_{i = 1}^n \sum_{j = 1}^m \frac{(O_{i,j} - E_{i,j})^2}{E_{i,j}},$$
with $O_{i,j}$ and $E_{i,j}$ being the observed and expected values respectively.  This sum combined with the number of degrees of freedom gives the chi-squared value for the data set.  Given the number of rows, $n$, and columns, $m$, the number of degrees of freedom is given by: $$df=(n-1)(m-1).$$

\subsection{Secure computation}

PICCO \cite{picco} is one of the available compilers for converting a user program which is intended to handle private data into its secure distributed implementation. It uses threshold linear secret sharing as the underlying mechanism for realizing secure computation or outsourcing. All participants are divided into three groups: input, computational, and output parties. Each input party secret-shares its input data among the computational parties, the computational parties securely carry out the computation on secret shared data, and the output is reconstructed by each output parties entitled to learn the result after receiving the shares of the result from the computational parties. There are no requirements on how these groups are composed with the exception that the number of computational parties needs to be three or larger and security relies on the assumption that a majority of the computational parties is honest.

Because secure computation is based in secret sharing, secret shares have a similar bitlength to that of the clear text data and local operations are considered to be free because of their efficiency. Complexity is then measured in the number of interactive operations and the number of rounds (sequential interactive operations). All arithmetic associated with private values takes place in a field. Addition and subtraction of secret-shared integers, as well as multiplication of a secret shared integer by a known integer are free, while multiplication of two secret shared integers requires a single interaction and is considered as the basic building block. The complier provides support for many integer and floating-point operations on private values and many other features of a general-purpose language. The compiler also provides multiple mechanisms for parallelizing the computation to improve the round complexity of securely executing a program, which often leads to improved performance. 

When PICCO is used to execute non-trivial user programs, informed users can rely on several features and aspects of the compiler to make their programs run more efficiently. In addition to the parallelization mechanisms mentioned above, Some of these methods include minimizing the use of expensive operations, restructuring the program to result in improved performance, and declaring variable that hold private data to be of the minimum necessary bitlength. We use these techniques to optimize two programs which feature statistical computations often utilized by institutional research groups and other potential research collaborators.

\section{Secure Computation of Statistical Tests}

Here we briefly describe how each of the two programs was designed.

\subsection{Standard deviation}

In our setting, the input data set comes from multiple parties and is privately read into an array for analysis. Because the amount of data can be quite large in practice, efficiency in processing is desirable. To achieve this, we implement the program to avoid costly operations when possible. In particular, we execute operations in parallel when possible, replace floating-point operations with integer operations, and minimize the number of divisions (which have a much higher cost than any other integer operations).

In our implementation, the overall data set is partitioned between different input sources. Thus, $N$ in equation \ref{eq:std-dev} is the sum of the individual data set sizes. The number of input parties can also vary and therefore the program begins by receiving the number of input parties and
the size of the input data corresponding to each party, after which the data is read into a private array. Our program itself can be found in Appendix~\ref{sec:sd-prog}. 

The computation then proceeds by determining the mean as the sum of all data items divided by the total number of data entries. Note that we use integer division and it is assumed that an integer mean value will provide enough precision as the data items are integers themselves. For each data item $x_i$ we consequently compute $(x_{i} - avg)^2$ in parallel, add the computed values and divide the sum by the total number of data entries. Finally, we want to compute the square root of this value to obtain the standard deviation of the data set. Note that in certain cases the square root computation can be avoided (e.g., when the standard deviation is always compared to another value or when directly the squared standard deviation is used), but for the purposes of this work we perform the overall computation.

The most interesting and non-trivial component of the standard deviation computation is the square root operation. Let the square root operation take integer $a$ as the input and output $b$ as an approximation for $\sqrt{a}$. We implement this function to utilize the better performance of integer over floating-point arithmetic as much as possible. As such, one parameter is the desired precision level $prec$, which will result in computing $prec$ decimal points of the output. In that operation, we first multiply the input value by $100^{prec}$ so that we will have the desired number of digits after taking the square root. Next, we follow the Newton-Raphson method of iterations using the following formula for the $k$th iteration:
$$b_k = \frac{1}{2} \left(b_{k-1} + \frac{a}{b_{k-1}}\right).$$
To reduce the number of divisions required, we used an equivalent rewriting of the formula as:
$$b_k = \frac{b_{k-1}^2 + a}{2b_{k-1}}.$$
Ordinarily, one would check the magnitude of error by computing $|a - b_k^2|$, but having this as a loop conditional in our setting would result in privacy leakage. If a computational party knows how many times the program looped before obtaining the desired error threshold, it can narrow down the possible value of the square root. Instead, we must rely on the bounded error rate of the Newton-Raphson method to know we have obtained our desired error rate without doing an unnecessarily large number of iterations.
To find the bound on the error at each iteration, we have that the error, $\epsilon,$ at the $k$th step is $\epsilon_k = |a - b_k^2|.$

To this point we have only used integers in the square root computation. The final step involves converting our integer result to a floating point result by dividing once by $10^{prec},$ offsetting our original multiplication by $100^{prec}$ prior to finding the square root. Thus, we have obtained the standard deviation to output to the appropriate parties.

\subsection{Chi-squared}
\label{sec:chisq}

In the chi-squared test, the input consists of a private matrix of size $n \times m$. We assume that each input party contributes a single row of the matrix and all input parties contribute data (population counts) corresponding to the same categories. All elements of the matrix including the sample size at each input party are considered private. (Note that in some applications the sample size contributed by each party is treated as public knowledge, in which case the secure computation can be carried out more efficiently. We, however, treat the general case when the population size at each contributing party remains private.)

Our program commences by reading the number of input parties $n$ and the number of data categories $m$ to be drawn from each input source, after which the data is read from the input parties in the secret-shared form.
We next iterate through the elements of the matrix to compute the sum of the elements in each row, in each column, and the overall sum of all elements of the matrix.

Most of the work comes from iterating over each cell and computing its chi-squared contribution. This is where we can significantly improve program execution time compared to the baseline version by optimizing the code. In particular, the formula used for computing the chi-squared contribution for each cell in an $n \times m$ data set is $$X_{i,j} = \left(x_{i,j} - \frac{ \sum_{k=0}^n x_{k,j} (\sum_{h=0}^{m} x_{i,h}) }{ \sum_{k=0}^{n} \sum_{h=0}^{m} x_{k,h}}\right)^2 / \left(\frac{(\sum_{k=0}^{n} x_{k,j})(\sum_{h=0}^{m} x_{i,h})}{\sum_{k=0}^{n} \sum_{h=0}^{m} x_{k,h}}\right),$$ where $x_i,j$ is the value stored at the $i$th row and $j$ column of the data matrix. Because we know the row, column, and overall sums, the above formula can be rewritten as $$X_{i,j} = \frac{(x_{i,j} - s_c s_r/s_o)^2}{s_c s_r/s_o},$$ where $s_r$, $s_c$, and $s_o$ denote the row, column, and the overall sum, respectively.
With preference given to integer computations over floating point due to cost, this would lead us to 1 integer multiplication followed by 4 floating-point operations (2 divisions, 1 multiplication, and 1 subtraction).  However, we can express the formula as $$X_{i,j} = \frac{(x_{i,j}s_o - s_c s_r)^2}{s_c s_r s_o},$$ 
in order to reduce the number of floating point-operations as well as the number of divisions.  This leaves us with only one floating-point operation to perform per cell.  All of $X_{i,j}$'s can be computed in batch, since cell computations are independent in this step.

Next, we need to sum these cell contributions which we achieve through utilizing batch operations again. While there are techniques for performing any number of integer additions in a constant number of rounds, these techniques do not generalize to floating-point additions. Thus, we perform batch additions in pairs in a tree-like fashion with the number of sequential additions equal to $\log_2(n \cdot m)$. This sum is the desired chi-squared statistic and is output to the appropriate parties. Our program can be found in Appendix~\ref{sec:cs-prog}.

\section{Performance Evaluation}

Here we present the timing results of running the two sample programs.  These timings were obtained by running each program fifteen times for shorter executions and five times for larger executions.  These results were then averaged to get the timings presented in the tables. All experiments were run by three computational parties, which are assumed to not collude with each other. The machines were 2.4 GHz desktops running Red Hat Linux and were connected through a 1Gb/s link.

\begin{table}[t]
\centering
\begin{tabular}{|c|c|c|c|c|c|c|c|c|} \hline
Input size & $2^4$ & $2^6$ & $2^8$ & $2^{10}$ & $2^{12}$ & $2^{14}$ & $2^{16}$ & $2^{18}$\\\hline
Standard deviation & 0.58 & 0.58 & 0.59 & 0.59 & 0.61 & 0.66 & 0.89 & 1.80 \\ \hline
\end{tabular}
\caption{\label{tab:StdDev}Execution times of the standard deviation program in seconds.}
\end{table}
Table~\ref{tab:StdDev} shows performance of the standard deviation program for different data sizes. The overall performance is dominated by the square root operation for small data sets and begins to exhibit linear dependency on the data set size for large sizes. 
We also ran a program with no optimizations except that square root computation was carried out on integers instead of floating-point numbers. 

In our experiments, there was no significant improvement in the timings of the square root computation compared to its straightforward version with a single optimization. This is due to the fact that the experiments used a low precision value (and the benefits of the optimizations are more pronounced with a larger number of algorithm's iterations) and the fact that implementation of the division operation in PICCO is optimized for some special cases (such as division by a (known or private) power of 2) to result in much faster performance than the general case.
The overall performance of the program, however, is substantially faster for the optimized version (which took more than 27 seconds for input of size $2^{18}$) because of the use of parallelism.

\begin{table}[t]
\centering
\begin{tabular}{|c|c|c|c|c|c|c|c|c|c|c|} \hline
Input Size & $4 \times 4$ & $4 \times 8$ & $4 \times 16$ & $4 \times 32$ & $4 \times 64$ & $4 \times 128$ & $8 \times 8$ & $8 \times 16$ & $8 \times 32$ & $8 \times 64$ \\\hline
Chi-squared & 0.38 & 0.59 & 0.93 & 1.62 & 2.94 & 5.59 & 0.96 & 1.64 & 2.97 & 5.61 \\\hline
\end{tabular}
\caption{\label{tab:ChiSq}Execution times of the chi-squared program in seconds.}
\end{table}
Table \ref{tab:ChiSq} shows performance of our optimized chi-squared program for different dimensions of the input matrix. 
The program uses parallelism and formula restructuring to reduce the number of floating-point operations as described in section~\ref{sec:chisq}. As a result, all run times are on the order of seconds or less. 
Our optimizations resulted in substantial improvements in the running
time compared to the unoptimized version. The unoptimized run times range between 2.8 and 89.6 seconds for the same data sizes with improvement ranging between 86\% (for small input sizes) to 94\% (for large input sizes). The run time improvement was measured above 90\% for all data sizes above $4 \times 4$. Because the improvement percentage increases along with the data size, the trend suggests that the reduction in run time will be even greater for larger data sizes if needed in practice. Because chi-squared tests are run on categorical data, the data sets used in practice are likely to be computationally feasible based on the timings of the data shown above.

\section*{Acknowledgments}

This work was supported in part by grants CNS-1223699 and CNS-1319090 from the National Science Foundation and FA9550-13-1-0066 from the Air Force Office of Scientific Research. Any opinions, findings, and conclusions or recommendations expressed in this publication are those of the authors and do not necessarily reflect the views of the funding agencies.

\bibliographystyle{plain}
\bibliography{refs}

\begin{thebibliography}{1}

\bibitem{Adams2012}
J.~Adams.
\newblock Collaborations: The rise of research networks.
\newblock {\em Nature}, 490(7420):335--336, October 2012.

\bibitem{eagan2014american}
K.~Eagan, J.~Lozano, S.~Hurtado, and M.~Case.
\newblock The american freshman: National norms fall 2014.
\newblock Cooperative Institutional Research Program at the Higher Education
  Research Institute, 2014.

\bibitem{yau2008}
S.~Yau and Y.~Yin.
\newblock A privacy preserving repository for data integration across data
  sharing services.
\newblock {\em IEEE Transactions on Services Computing}, 1(3):130--140, July
  2008.

\bibitem{picco}
Y.~Zhang, A.~Steele, and M.~Blanton.
\newblock {PICCO}: {A} general-purpose compiler for private distributed
  computation.
\newblock In {\em ACM Conference on Computer and Communications Security
  (CCS)}, pages 813--826, 2013.

\end{thebibliography}

\appendix
\section{User Programs}

\subsection{Standard deviation program StdDev} 
\label{sec:sd-prog}

{\tt \small
\begin{lstlisting}[breaklines=true]
public void sqrRoot(private int num, public int prec){
	private int k = 1;
	public int i = 0;
	for(i = 0; i < prec; i++){
		num = num * 100;
	}
	for(i = 0; i < prec + 10; i++){
		k = (k*k + num) / (2*k);
	}
	private float<32,9> ans;
	ans = (private float<32,9>) k;
	private int temp = 1;
	for(i = 0; i < prec; i++){
		temp = temp * 10;
	}
	private float<32,9> tempFloat;
	tempFloat = (float<32,9>)temp;
	ans = ans / tempFloat;
	smcoutput(ans, 1);
}

public int main() {
	public int numParties, i;
	smcinput(numParties, 1);
	public int sizes[numParties], spots[numParties];
	smcinput(sizes, 1, 512);
	
	public int total = 0, maxSize = 0;
	for(i = 0; i < numParties; i++){
		if(sizes[i] > maxSize){
			maxSize = sizes[i];
		}
		spots[i] = total;
		total += sizes[i];
	}
	private int all[total], temp[numParties][maxSize];

	smcinput(temp, 1, 512, 512);
	for(i = 0; i < numParties; i++){
		for(public int j = 0; j < sizes[i]; j++){
			all[spots[i] + j] = temp[i][j];
		}
	}
//Now work with dataset
	private int mean = 0;
	for(i = 0; i < total; i++){
		mean += all[i];
	}
	private int tempTotal = total;
	mean = mean / tempTotal;
	private int stdDev = 0;
	for(i = 0; i < total; i++)[
		all[i] = (all[i] - mean) * (all[i] - mean);
	]
	for(i = 0; i < total; i++){
		stdDev += all[i];
	}
	stdDev = stdDev / tempTotal;
	sqrRoot(stdDev,1);
}
\end{lstlisting}
}

\subsection{Chi-squared program ChiSq}
\label{sec:cs-prog}

{\tt \small
\begin{lstlisting}[breaklines=true]
public int main() {
	public int numParties, size, i, j;
	smcinput(numParties, 1);
	smcinput(size, 1);
	
	public int sizepow;
	private int temp[numParties][size];
	smcinput(temp, 1, 8, 64);
	smcinput(sizepow, 1);	

//Now work with dataset
	private int partyTotal[numParties];
	private int typeTotal[size], total = 0;
	for(i = 0; i < size; i++){
		typeTotal[i] = 0;
	}
	for(i = 0; i < numParties; i++){
		partyTotal[i] = 0;
		for(public int j = 0; j < size; j++){
			partyTotal[i] += temp[i][j];
			typeTotal[j] += temp[i][j];
			total += temp[i][j];
		}
	}
	
	private int temp1[numParties][size], temp2;
	private float<32,9> answer = 0.0, tempFloat1[numParties][size], tempFloat2[numParties][size];
	public int totalSpots = numParties*size;
	private float<32,9> tempAnswer[totalSpots];
	for(i = 0; i < numParties; i++)[
		for(j = 0; j < size; j++)[
			temp1[i][j] = typeTotal[j] * partyTotal[i];
			temp[i][j] = temp[i][j] * total;
			temp[i][j] = temp[i][j] - temp1[i][j];
			temp[i][j] = temp[i][j] * temp[i][j];
			temp1[i][j] = temp1[i][j] * total;
			tempFloat1[i][j] = (private float<32,9>)temp1[i][j];
			tempFloat2[i][j] = (private float<32,9>)temp[i][j];
			tempAnswer[i*size + j] = tempFloat2[i][j] / tempFloat1[i][j];
		]
	]
	for(i = 0; i < sizepow; i++){
		for(j = 0; j < totalSpots/2; j++)[
			tempAnswer[j] = tempAnswer[j] + tempAnswer[totalSpots - j-1];

		]
		totalSpots /= 2;
	}
	answer = tempAnswer[0];
	smcoutput(answer, 1);
}
\end{lstlisting}
}

\end{document}